# Role of sonication pre-treatment and cation valence in nano-cellulose suspensions sol-gel transition

C. A. Maestri[a], M. Abrami[b], S. Hazan[c], E. Chistè[a], Y. Golan[c,d], J. Rohrer[e], A. Bernkop-Schnürch[e], M. Grassi[b], M. Scarpa[a], P. Bettotti[a]*

[a] Nanoscience Laboratory, Department of Physics, University of Trento, Via Sommarive 14, 38123, Povo (TN), Italy
[b] Department of Engineering and Architecture, University of Trieste, Piazzale Europa 1, 34127 Trieste, Italy
[c] Ilse Katz Institute for Nanoscale, Science and Technology, Ben Gurion University of the Negev Beer Sheva 84105, Israel
[d] Department of Materials Engineering, Ben Gurion University of the Negev, Beer Sheva 84105, Israel
[e] Department of Pharmaceutical Technology, Institute of Pharmacy, University of Innsbruck, Innrain 80/82, Innsbruck, Austria
*Corresponding author: paolo.bettotti@unitn.it

**ABSTRACT**
Sol-gel transition of carboxylated cellulose nanocrystals is investigated using rheology, SAXS, NMR and optical spectroscopies to unveil the distinctive roles of ultrasounds treatment and ions addition. Besides cellulose fibers fragmentation, sonication treatment induces fast gelling of the solution. Gelation is induced independently on the addition of cations, while the final rheological properties are highly influenced by the type, the concentration as well as on the sequence of the operations since salts must be added before sonication to produce stiff gels.
Cations with various charge and dimension have been associated to ultrasounds to induce gelation and the gel elastic modulus increase proportionally with the charge over the ion size ratio. SAXS analysis of the $Na^+$ hydrogel and $Ca^{2+}$ hydrogel to which the ion was added after sonication shows the presence of structurally ordered domains where water is confined as indicated by 1H-NMR investigation of the dynamic of water exchange in the hydrogels. Conversely, separated phases containing essentially free water, characterize the hydrogels obtained by sonication after $Ca^{2+}$ addition, confirming that this ion induces irreversible fiber aggregation. The rheological properties of the hydrogels depend on the duration of the ultrasound treatment and it enables the design of materials programmed with tailored energy dissipation response.

**INTRODUCTION**
Nanocellulose (NC) is a renewable and biocompatible material with interesting and versatile properties which allow its integration in a huge number of applications, as has been extensively reviewed [1,2]. The procedures to break natural cellulose and obtain nano-sized structures are usually based on the combination of chemical modification or enzymatic hydrolysis with mechanical refinement [3,4]. Fine changes of these procedures give rise to different nanostructure morphology: branched nanofibrils with amorphous regions (CNF) and rod-like rigid nanocrystals (CNC) [5]. TEMPO-mediated oxidation of cellulose followed by sonication provides well dispersed, negatively charged CNC [6].
Despite the large interest on NC and its applications, several basic aspects regulating NC properties and its interaction with the environment are still unclear. Concerning the structure investigation, the effort has been addressed mainly toward the understanding of the liquid crystalline self-assembly resulting in ordered helical structures with peculiar mechanical and optical properties [7-9]. The self-assembly of NC or NC-composites into soft hydrogels [10,11] has been characterized in terms of macroscopic parameters such as mesh size, charge density, gelation rate, mechanical performances, or stability [12-15]. In this context, rheology experiments have been performed and a gel-like behavior of NC suspensions with an elastic response even at a low concentration [16] has been reported. In general, the rheological behaviour of NC suspensions is strongly dependent on NC production: mechanical fibrillation without chemical modification produces suspensions with flocculated structure while NC which underwent chemical processes, produces suspensions with better colloidal stability [17]. The static and dynamic rheological behavior of hydrogels of rod-like CNC suggest that liquid crystal domains consisting of self-organized ordered structures are present in CNC [16]. Divalent or trivalent cations ($Ca^{2+}$, $Zn^{2+}$, $Cu^{2+}$, $Al^{3+}$, and $Fe^{3+}$) induce gelation of negatively charged CNF and form interconnected porous nanofibril networks [18]. Dynamic viscoelastic measurements performed on these gels and SEM images measured on dried samples [19] reveal storage moduli and a mesh size strongly related to the valence of the metal cations and their binding strength with carboxylate groups on CNF obtained by TEMPO-mediated oxidation. Though these results shed light on the role of the cross-linking reactions and electrostatic interactions in the formation of three-dimensional NC networks, little is known about the contribution of hydrogen bonds. For example, only recently it has been proposed a mechanism relating the role of covalent and hydrogen bonds to explain cellulose elastic properties [20] and the exploitation of recoverable physical bonds as sacrificial bonds for energy dissipation has been suggested to reduce internal hydrogel damage under stress and increase fatigue resistance [21]. Hydrogels contain huge amounts of water which is an excellent competitor for intra-and inter- fibril hydrogen bonds and polysaccharides are the biomolecules of excellence for the formation of hydrogen bonds. In this regard NC behaves as a typical polysaccharide and its structure and dynamics in solution strongly depends on electrostatic bonds and on the surface available for their occurrence. Different dynamic regimes of the water molecules have been observed within and on the surface of polysaccharide-based or synthetic hydrogels: free interstitial water which does not take part in hydrogen bonds with hydrogel molecules; bound water, which is directly bound to the chains and semi-bound water, with intermediate properties [22,23]. As far as we know the presence and the dynamic behaviour of these water molecules has not been investigated in NC hydrogels which could substantially differ from more traditional hydrogels because formed by the assembly of nanostructures rather than polymer chains. In this work the mechanical and chemical sol-gel transition of CNC have been investigated by the combined use of spectroscopic and rheological techniques. We found that rod-like CNC undergo a sol-gel transition process apparently similar to that observed for flexible polymer chains and they form stable hydrogels containing rigid structural domains inside which water molecules are confined. Moreover the stability of the hydrogel depends on the sonication treatment and this fact foreseen the possibility to fabricate programmable gels that behaves differently depending on the amount of energy they have to dissipate.

**RESULTS**
Sol-gel transition in TEMPO oxidized cellulose (TOC) nanocrystals aqueous solutions:
40 mL of aqueous slurry containing 6 $mgmL^{-1}$ TOC at pH 7 was sonicated for variable times. Before sonication the solution was highly heterogeneous and formed by macroscopic aggregates of fibers. The turbid and flocculent suspension became progressively a homogeneous and viscous jelly solution with the sonication. After 120 s of sonication, the gel was uniform and transparent. Accordingly, the transmittance % of the TOC suspensions reached a plateau value in the range 120-240 sonication time (see Fig. 1(a)). Investigation of morphology changes of the TOC nanostructures with sonication showed that the branched TOC fibrils (visible in Supplementary Fig. S1 which is the optical microscopy image of TOC after 30 s sonication) turned into rod-like crystallites (shown in Supplementary Fig. S1 which is the AFM image obtained after 480 s sonication) and this transition was almost complete after 120 s sonication time. The dispersion of the fibers is driven by their polyelectrolytes nature. In fact at pH 7 the TOC nanocrystals bear an average negative charge on the 29% of the cellobiose units, as determined by conductometric titrations.

The dynamic rheological behaviour of 6 mg/ml TOC suspensions obtained by frequency sweep tests confirms the results of the optical properties and of the structural evolution of the fibres. Indeed, frequency sweep tests indicate a gel-like behaviour of the samples sonicated for 30-360 sec, being G' > G'', and both G' and G'' roughly independent from the frequency. Figure 1(b) presents the dependence of G' (conventionally evaluated at 1 Hz) versus sonication time. It shows that gel strength (i.e. its elastic component) increases up to 120 s and then decreases, suggesting a reduction of the interaction points (crosslinks) among different TOC rods. However, the typical gel behaviour is always attained whatever the sonication time.

Transversal nuclear magnetic relaxation rate ($T_2^{-1}$) of water is used to investigate the state and dynamics of water in the TOC suspensions and hydrogels at different sonication times. Multi-exponential analysis of the magnetization decay reveals that for short sonication times (shaded region in Fig. 1(a), for sonication time up to 120 s), two relaxation times are required to describe relaxation of the magnetization. This fact suggests the coexistence of two different proton environments: macroscopic dispersed fibres correlate with the slower relaxation rate (more similar to that of free water molecules), while homogeneously, randomly arranged nanocrystals are compatible with the faster rates. Once the gel is completely formed, a single time component is sufficient to describe the system, thus supporting the idea of a homogeneous structure of the material. Moreover the average relaxation times ($T_{2m}$) decreases with the sonication time, indicating that fibers progressively detach from each other so that the polymeric surface available for the interaction with water molecules increases (data about $T_{2m}$ are reported in Supplementary Fig. S2).

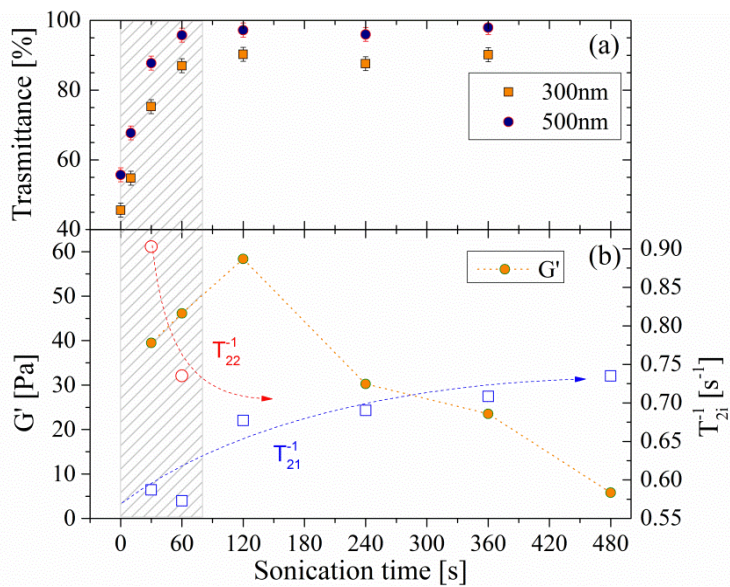

Fig. 1 (a) Transmittance of TOC vs. sonication time. Square orange points refer to UV wavelengths (300 nm), while blue round points indicate visible (500 nm) wavelength. (b) Evolution of both elastic modulus (filled circles, left axis) and of relaxation rates (open symbols, right axis). Arrows are guides for the eyes. The shaded region underlines the sonication times during which gel structure forms.

Despite their gel nature, the connectivity within these gels is not very high, probably because of the high rigidity and limited length of TOC nanofibers, which forbid the formation of highly entangled structures. In fact, the flow curves show that, for sonication times lying in the range 30 – 360 s, a sudden drop of viscosity happens at stress around 20-60 Pa (see Supplementary Fig. S3) and indicates a fracture of the internal structure. Indeed, assuming that the shear modulus of the gels is represented by the average *G'* value over the frequency range explored and that Flory theory holds [24], the resulting average mesh size is wide (≈ 50 – 70 nm). This datum indicates the presence of a tenuous, transient network (statistical network), similar to that observed for weak polysaccharide gels [25,26]. Despite these considerations, the effect of sonication is not negligible as, non-sonicated TOC suspension showed *G'*< *G''*, *G'* frequency dependent, low and constant viscosity over the whole interval of stress values.
The gel shows a reversible behavior and, upon long sonication time (> 360 s), it tends to revert to a viscous solution as reported by the rheological characterization (G' as well as viscosity decrease).
The network structure formed by the TOC crystallites was investigated by SAXS and the typical profile of aqueous solutions was found, irrespectively of the sonication time.

The overview of the experimental behavior of the aqueous suspensions of TOC indicates that 30-120 s sonication breaks the fibrils in correspondence of the amorphous regions producing isolated TOC crystallites which form a weak physical gel. By further sonication a solution state is reached, probably because the supplied energy breaks the hydrogen bridges between the TOCs. The role of hydrogen bonds is supported also by the dynamics of water exchange in TOC suspensions. At 30 s sonication time water is in fast exchange between the TOC fibrils surface and the bulk, being the bulk contribution predominant since the $T_{21}^{-1}$ is similar to that of water. Sonication lasting 30-120 s confines the water in either the TOC domains (with mesh size in the order of 50 nm) or in its "bulk" phase. As a consequence, the exchange rate between these two environments is slow. Further sonication uniformly disperses the TOCs crystallites in the bulk and makes more binding sites available for hydrogen bonds with water itself. In this case fast exchange regime between free and bound water holds, the transversal magnetization decay is fitted to a single exponential and the correlation time of bound water determines the increase of the measured relaxation rate.

Sol-gel transition in salt-added TOC solutions:
Similarly to other polyelectrolyte systems (e.g. alginates [27]), cations increase the interactions among elementary elements that form the gel and drastically modify the gel properties. Despite TOC hydrogels are not formed by long, intertwined, polymeric chains, they share several typical characteristics of these systems and cations easily induce gelation in TOC solutions. In our case sonication to disperse TOC, followed by salt addition, produces homogeneous hydrogels; conversely, sonication performed on TOC solutions containing 100 mM of multivalent cations produces a visually inhomogeneous hydrogel where compact macroscopic structures are surrounded by a waterish suspension (detailed flow curves for these cases are reported in Supplementary Fig. S4).

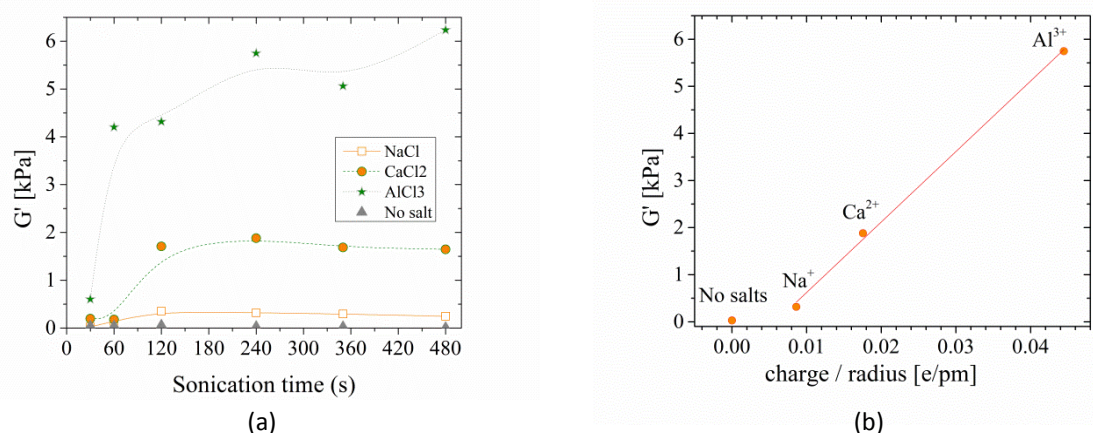

(a) (b)
Figure 2 (a) Storage modulus measured @ 1Hz for gels formed using 100mM of different cations versus sonication time. Lines are guides for the eyes. (b) G' vs the ratio of cation valence over cation radius for samples sonicated 240 s. The red line is the linear fit of the data.

Mechanical properties increase proportionally to cations valence and to their concentration. Figure 2(a) reports how G' (conventionally evaluated at 1 Hz) changes upon gelation for 100 mM solutions of different cations. We found that pre-treatment of at least 120 s is required to form a homogeneous gel with a nearly stabilized G' value.
Gel stiffness increases proportionally to cation valence. Figure 2(b) reports the value of G' for solutions sonicated 240 s versus the ratio of cation valence over cations radius (assuming 116 pm $Na^+$, 114 pm for $Ca^{2+}$ and 67.5 pm for $Al^{3+}$). G' is found to increase linearly ($\chi^2$=0.999) for all the samples added with salts. We did not check higher valences because of the marked different reactivity of such cations that might introduce significant differences in the gelation mechanism. The linear trend is typical also of polymeric systems [28] and the main difference is the reduced G' modulus achievable with TOC nanofibers, due to their limited capability to entangle.
To note that the small G' value obtained using $Na^+$ permits the formation of a homogeneous gel irrespectively of the order of sonication and salt addition operation. This suggests that the coordination of $Na^+$ by TEMPO oxidized cellulose is not strong enough to contrast the breaking of sacrificial bonds induced by ultrasonication.
Assuming that Flory theory [24] to holds also for TOC-based gels, we estimate the average mesh size assuming that the shear modulus is the average *G'* value over the frequency range explored. By this calculation we can estimate an average mesh size of about 60, 25, 15 and 10 nm for no salt addition, 100 mM NaCl addition, 100 mM $CaCl_2$ addition and 100 mM $AlCl_3$ addition, respectively, after 240 s sonication. As expected, mesh size is inversely proportional to G' module of the gels.

We noticed that upon gelation (that is for sonication times longer than 120 s), $T_{21}^{-1}$ component assumes comparable value for both NaCl and $CaCl_2$ gels and it remains constant irrespectively of the sonication time. $T_{22}^{-1}$ is also constant vs sonication time but it assumes different values for the gels produced with either the mono- and bi-valent cations. The suspension with no added salts shows only one component. These data are summarized in Table 2. Thus the value of $T_{22}^{-1}$ component is correlated with the G' modulus of the gel (as shown also in Supplementary Fig. S5) and indicates a strong dependence of the elastic properties of the hydrogel from its local structure mediated by both the presence of cations as well as hydrogen interactions.

Table 2.

| Salt | $T_{21}^{-1}$ (s$^{-1}$) | $T_{22}^{-1}$ (s$^{-1}$) |
|---|---|---|
| NaCl[a)] | 0.57 ± 0.03 | 0.88 ± 0.13 |
| CaCl$_2$[b)] | 0.59 ± 0.06 | 1.13 ± 0.06 |
| No Salt | 0.69 ± 0.03 | - |

[a), b)]100 mM salt was added to suspensions of 6 mgmL$^{-1}$ TOC sonicated for 240 s.

Because $T_{22}^{-1}$ components do not change with sonication time, we are allowed to compare their relative amplitude as a function of the sonication treatment, as reported in Fig. 3.

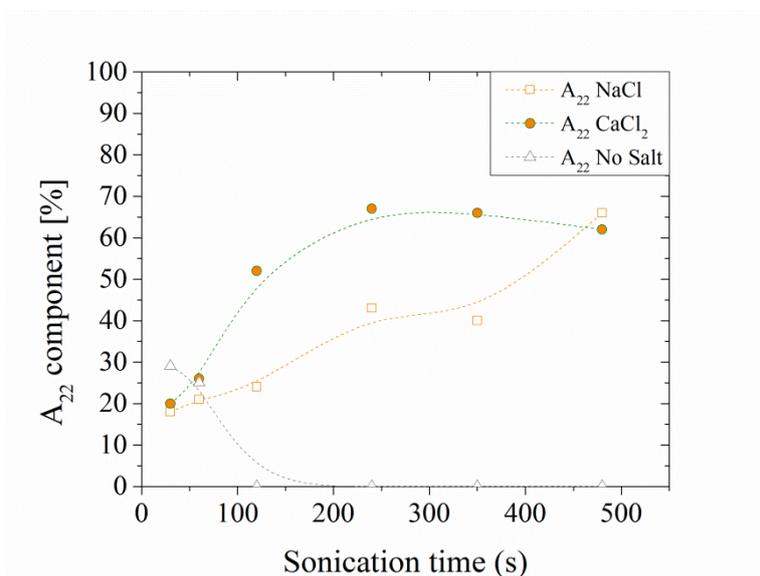

Figure 3. Amplitudes of the A$_{22}$ components plotted for NaCl (squares), CaCl$_2$ (circles) and solution without added salts (triangles).

The $A_{22}^{Ca^{2+}}$ component saturates for sonication times longer than 120 s, suggesting that nanofibrils arrangement reaches a stable configuration. On the other hand, $A_{22}^{Na^+}$ steadily increases supporting the idea that the structure of the Na-induced gel is weak and dependent on the initial conditions (that is the electrostatic interaction of Na$^+$ is not enough to structure the gel). As pointed out before, the $A_{22}^{sol}$ component of pure solutions disappear for sonication times longer than 120 s.

The hydrogels obtained after addition of NaCl, CaCl$_2$ and AlCl$_3$ to TOC suspensions sonicated for 240 s were investigated by SAXS measurements. Typical spectra are reported in Fig 4.

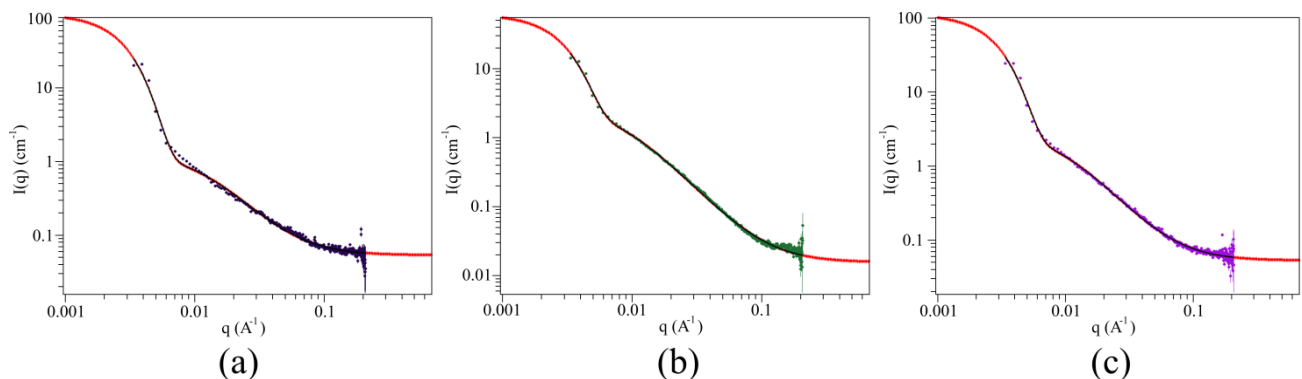

(a) (b) (c)

Figure 4 SAXS profile of sonicated TOC suspensions after salt addition. Suspensions of 6 mgmL$^{-1}$ TOC have been sonicated for 240 s then 100 mM NaCl(a), CaCl$_2$(b) and AlCl$_3$(c) were added. The dots are the experimental points, the red line is the fit of the experimental points with the Gauss-Lorentz Gel Model (see Eq. 2 in Sec. Methods).

SAXS analysis does not reveal any peak indicating the presence of structurally ordered domains. The SAXS profiles of the hydrogels (obtained adding 100 mM NaCl CaCl$_2$ and AlCl$_3$ to TOC solutions sonicated for 240 s) were properly fitted with the Lorentz-Gauss models of Eq. 1 and point out the presence of two structural length scales, those values are reported in Table 3. The different length scale could be associated to higher density domains (solid-like) and lower density regions as reported for gel-like polymeric networks [29]. The static correlation length is larger than the dynamic one and corresponds to the correlation within long-lived entanglements or their average size compared to the dynamic correlation length corresponding to the distance to which the movement of the flexible polymer chain is correlated [30].

On the contrary of findings reported by other authors [31], our samples do not show any ordered domain. This difference is probably due to thoroughly washing of the sample we use to neutralize the solution, instead of adding HCl. This fact modifies the ionic strength of the solution and reflects in material with different properties (e.g. we noticed a highly increased gas barrier property of cellulose films obtained with this method [32]).

TABLE 3 Static and dynamic correlation lengths of TOC hydrogels formed by the addition of different salts (100 mM) after 240 s sonication.

| Salt (100 mM) | Static length (Å) | Dynamic length (Å) | Mesh size (Å)[d] |
|---|---|---|---|
| NaCl | 416 ± 144[c] | 77 ± 7[c] | 250 |
| CaCl$_2$ | 474 ± 28[a] | 127 ± 11[a] | 150 |
| AlCl$_3$ | 469 ± 51[b] | 106 ± 13[b] | 100 |

[a] average and standard deviation (s.d.) of 3 different samples; [b] average and s.d. of 5 different samples; [c] average and s.d. of 4 different samples; [d] values estimated assuming Flory theory.

Notably, both static and dynamic correlation lengths are independent on the crosslinking cation and their values do not easily correlate with the characteristic nanocrystal size (about 300 x 5 x 5nm). The static correlation length might describe the typical size of high density regions formed by the overlapping CNC. Under these assumptions, the density of CNC would be constant across all samples (as confirmed by both optical and NMR analysis), irrespectively from the cations. On the other hand, the dynamic correlation length depends on the local gel structure as it is affected by the strength of the crosslink created by the different cations. Thus, it directly correlates with the G' of the gel. The last column of Table 3 reports the values of the mesh sizes as estimated from the Flory theory. While there is a large discrepancy between SAXS and Flory theory predictions for gels formed using NaCl; the values reasonably agree for gels formed using multivalent cations. This fact support the intuitive idea that rigid materials have structure made by small meshes (differently from that data recently reported in [31] for similar gels).

The comparison of gel formed by adding 10 mM and 100 mM concentrated solutions is discussed below. Figure 5 reports the frequency sweep tests for the gels obtained using different cations.

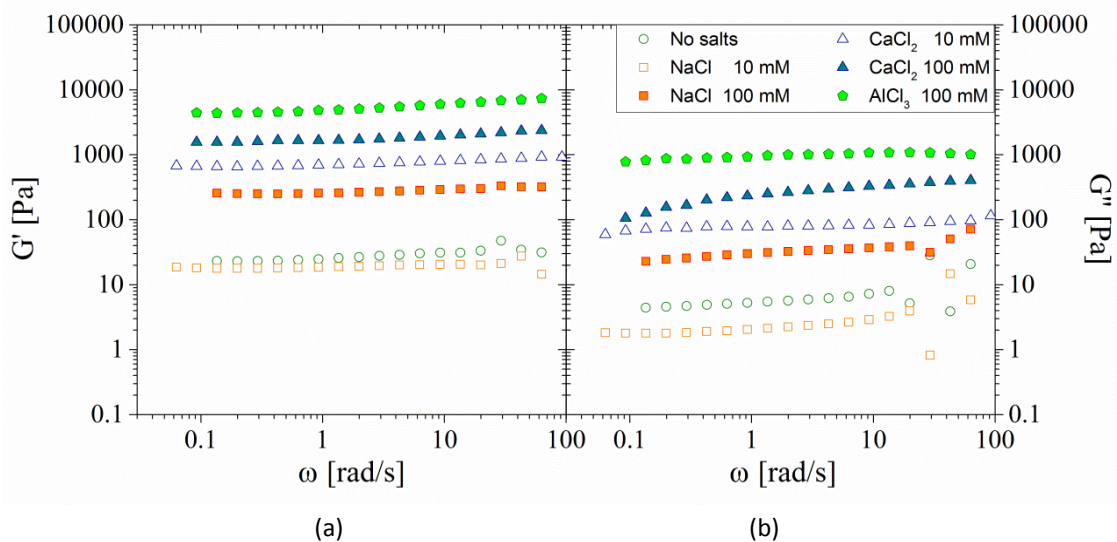

(a)  (b)

Figure 5. Frequency sweep tests of TOC gels created adding different cations and concentrations to suspensions of 6 mgmL$^{-1}$ TOC sonicated for 240 s. (a) Elastic moduli; (b) viscous moduli.

While the cation valence has a profound influence on the gel stiffness, for a given cation, modest changes appear between the 10 mM and the 100 mM concentrations. The relatively small difference between the 10 mM and 100 mM concentrations might be due to a saturation of the TOC sites available for crosslinks.

The maximum value of G' depends on the sonication pre-treatment. Depending on if and what cation is added, G' can either increase or decrease with the sonication time. This fact suggests that gels are in metastable state. This hypothesis is confirmed by looking at the dynamics of the G' change vs sonication time. We assume the G' value at 120 s of sonication ($G'_{120}$) to be the highest achievable in each gel and we normalize the decrease of G' for different sonication pre-treatment against the $G'_{120}$ value: $\Delta G' = \frac{G'_{120} - G'}{G'_{120}} x\, 100$

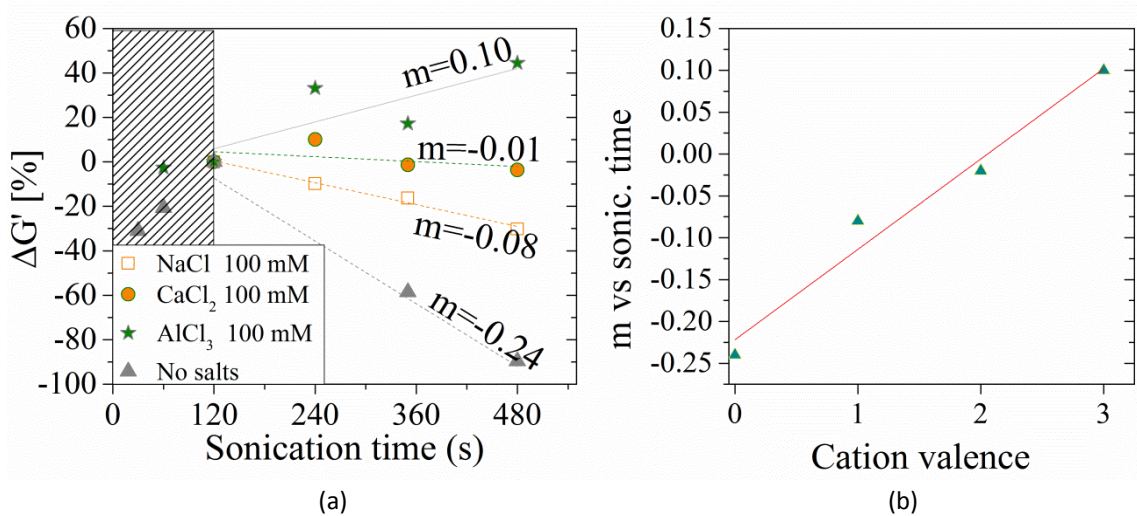

(a)  (b)

Figure 6. (a) Variation of G' (conventionally evaluated at 1 Hz) vs sonication treatment for different cations (the data are the same reported in Fig. 2). (b) Plot of the slope of G'(1 Hz) variation versus cation valence.

The result is that the ΔG' scales inversely proportional to the valence of the crosslinking cations. That is, given a cation valence, the pretreatment duration to bring the gel into the most stable state, scales proportionally to the valence. This evidence correlates with the trend of the amplitude of the $T_{22}$ magnetization decay that steadily increases in NaCl systems, while saturates in $CaCl_2$ for sonication times longer than 120 s. This result supports those reported in Fig. 1: a prolonged sonication treatment reduces the density of crosslinking points and, in turn, the G'. Such rearrangement is hindered in stiffer gels that require longer sonication to achieve their maximum G' and show a slower decrease (if any) of their Young's modulus.

**DISCUSSION**

We investigated the effect of the parameters which strongly affect the assembly of TOC in gel-like supramolecular networks: the energy released by sonication and the addition of small cations.

The overall analysis of the parameters highlights the following fundamental results:

1) TOC crystallites which are characterized by a rod-like, rigid structure, undergo a sol-gel transition process apparently similar to that observed for flexible polymer chains. This transition can be induced by non-exhaustive sonication, even though a clearer gel-like behavior is obtained if sonication is followed by salt addition;

2) the gelation mechanism is cation-dependent and mediated by hydrogen interactions;

3) the stability of the gel depends on the duration of the sonication treatment.

Compared to truly polymeric systems formed by long, entangled molecular chains, TOC short and rigid crystallites yet form gels thanks to the strong electrostatic interactions generated by hydrogen bonds and between cations and carboxyl groups. Gelation is nearly instantaneous and it started by simply hand shaking a sonicated solution added with cations. The fact that homogeneous gels are formed only if salts are added to fibres well dispersed by sonication suggests that strong interactions take place between the cations and the non-sonicated fibres that form inhomogeneous macroscopic aggregates. These aggregates limit the efficacy of the sonication treatment, contrast the dispersion of isolated nanocrystal and hinder the formation of a homogeneous gel phase.

These hypotheses are confirmed by the different behaviour of the gels formed by adding either mono- or multi-valent cations: softer gels (NaCl) have wider meshes that reduce their G' modulus. Moreover, despite the weaker interactions produced by $Na^+$, the flow curves of these gels shows a drop of the viscosity larger than the corresponding for the inhomogeneous gel formed with $Ca^{2+}$: the weaker interactions created by monovalent cations permit them to disperse homogeneously in the gel phase independently on the order of salt addition and sonication. On the other hand both $CaCl_2$ and $AlCl_3$-added gels form homogeneous phases only if cations are free to diffuse around isolated nanocrystals, aka only if TOC fibres are already well dispersed by the sonication. In the case of $Ca^{2+}$, the flow curves clearly show a higher threshold value for the viscosity drop compared to the $Na^+$-based gels.

The homogeneity of gels formed by TOC crystallites is confirmed by both optical and NMR analysis: gel transparency is very high and essentially no scattering appears in transmittance measurements, while NMR reveals nearly constant $T_{21}^{-1}$ rates for all gels, while the $T_{22}^{-1}$ parameter scales with the G' modulus and with the duration of the sonication treatment. SAXS analysis has unveil that all gels contain high-density domains of about 40-50 nm. Water confined in these domains exchanges slowly with the less structured regions, giving rise to a compartmentalized relaxation behavior. These high-density domains are independent of the added cation (i.e. similar dimensions and water relaxation rate were obtained). However the elasticity of the network is cation dependent, as shown in Fig. 2 and $Ca^{2+}$ than $Na^+$ is more effective in forming such rigid domains. This fact is confirmed by the higher value of the amplitude of the magnetization component relaxing with rate $T_{22}^{-1}$ which is ascribable to the overall number of water molecules confined in the rigid domains. This amplitude reaches its maximum value (about 70%) for $Ca^{2+}$ added to TOC sonicated for more than 120 s. Conversely, if $Na^+$ is added longer sonication times are required for the same amount of confined water. These results suggest that beyond the electrostatic interaction, the coordination by carboxylated groups contributes to the $Ca^{2+}$ effect, and are in line with the interaction models of cations with nanocellulose fibres proposed by [33] and with the stability of metal carboxylate complexes measured by [34]. Conversely, non-specific electrostatic interactions seem to predominate in the case of $Na^+$ which necessitates of more extensive sonication which breaks the hydrogen bonds between the TOCs and forms weaker gels. Unfortunately, similar conclusions cannot be extended to the case of $Al^{3+}$ because it affects the magnetization decay. We suppose the different behaviour of the $Al^{3+}$ to be related with a faster gelation dynamics induced by trivalent cation that produces inhomogeneous gels structures. Probably a more homogeneous dispersion of the cations across the gel volume would achieve a more stable state (stiffer gel). Such hypothesis is indirectly supported by the data reported in Fig. 6 in which $Al^{3+}$ is the only cation who does not achieve the maximum G' upon 120 s of sonication.

These results demonstrate that, despite their non-polymeric and rigid nature, TOC nanocrystals shows sol-gel transition similarly to polymers and produce gels with good mechanical properties. Moreover the short length of the TOC nanocrystals (of the order of few hundreds nm) enables a very fast sol-gel transition (few seconds) and the mechanical properties of the gels can be tuned by playing with both cation valence and sonication pre-treatment. Despite the negligible entanglement shown by the CNC and the fast initial stage of gelation induced by multivalent cations, reaching the maximum G' -for a given preparation condition- shows a complex and rich behaviour as it requires long sonication times. The large amount of energy required to bring the system into a stable state is probably connected with a hindered mobility of the CNC after the initial stage of the gelation; while the decrease of the G' by prolonged sonication treatments, might be due to reduction of crosslinking points of the nanofibers that, by reducing their entanglement, increases the mobility of neighbour fibres and decreases the overall gel stiffness. This behaviour might be used to design hydrogels programmed to behave differently depending on the amount of energy they have to dissipate.

**METHODS**

Reagents: All reagents used were from Sigma (St. Louis, MO) and were used without further purification. Never dried soft bleach pulp (Celeste90©) were received from SCA-Ostrand© (Sweden).

Preparation of Nanocellulose: 2,2,6,6-tetramethylpiperidinyloxy (TEMPO) oxidized cellulose (TOC) was obtained by a slightly modifying the procedure reported in [35]: 1 g of cellulose pulp was swollen in 100 mL water under stirring for 1 hr. Then 0.1 g NaBr, 1.75mL NaClO and 16.2mg TEMPO were added under vigorous stirring. The pH of the solution was maintained in the interval 10.5-11.0 by addition of 1 M NaOH until it remained almost constant (around 0.75mL NaOH were added). Then, the slurry of TOC was carefully purified by other chemicals and brought to pH 7 by repeated washings in ultrapure water and concentrated by a rotary evaporator (Heidolph, Schwaback Germany) to obtain a final TOC concentration of 6 mgmL$^{-1}$.

The sonication was performed by using an ultrasonic homogenizer (HD2200 Bandelin Sonoplus, Berlin Germany) equipped with a 13 mm titanium tip. An output power $W_{eff}$ of 160W was delivered in 40 mL of TOC slurry for a variable time.

Rheology experiments: rheological measurements were performed by a stress controlled rotational rheometer (Haake Mars Rheometer, 379-0200 Thermo Electron GmbH, Karlsruhe, Germany) equipped by parallel plate geometry (C35/1°, ϕ= 35 mm). The gap was fixed at 0.5 mm. The linear viscoelastic region was determined by stress sweep tests (in the range 0.01-500 Pa) keeping a constant frequency at 1 Hz. Frequency sweep tests were performed in the frequency range 0.01-10 Hz at a constant shear stress of 1 Pa (within the linear viscoelastic field) and the temperature was always set to 25 ± 1°C.

Nuclear Magnetic Resonance (NMR) measurements: the dynamics of water in the hydrogels was investigated by low-field 1H-NMR. To this purpose, the water protons transverse relaxation time ($T_2$) of the hydrogels and of the NC suspensions was measured at 25°C by a Bruker Minispec mq20 operating at 20.1 MHz (Karlsruhe, Germany). The CPMG (Carr–Purcell–Meiboom–Gill) sequence {90°[-τ-180°-τ(echo)]$_n$-$T_R$} with a 8.36 μs wide 90°pulse, τ = 250 μs and $T_R$ (sequences repetition rate) equal to 5 s was used. The criterion adopted to choose $n$ consisted in ensuring that the final FID intensity were about 2% of the initial FID intensity (in the light of this acquisition strategy, we verified that it was un-necessary adopting $T_R$ > 5 s). Accordingly, $n$ was approximatively equal to 700. Finally, each FID decay, composed by $n$ points, was repeated 36 times (number of scans). The relaxation times distribution ($A_i$, $T_{2i}$) was determined by fitting the FID time decay ($I_s$), related to the extinction of the *x–y* component of the magnetization vector ($M_{xy}$), according to its theoretical estimation $I$(t):

$$I(t) = \sum_{i=1}^{m} A_i e^{-\frac{t}{T_{2i}}} \qquad (1)$$

where t is time, $A_i$ is the amplitude of the magnetization of the protons relaxing with time $T_{2i}$. The number of exponential m was determined by minimizing the product $\chi^2*(2m)$, where $\chi^2$ is the sum of the squared errors and 2m represents the number of fitting parameters of Eq.(1) [36].

Microscopy measurements: a digital Atomic Force Microscope (NT-MDT Universal SPM scanning head SMENA) equipped with a Si tip, operating in semi-contact mode was used to investigate the morphology, size and size distribution of the TOC fibers. TOC solutions were suitably diluted with water to obtain isolated fibers. A drop of this solution was deposited on a silicon support and dried at 60°C in an oven. Image analysis was performed using the software Gwyddion v2.46 [37].

Small angle X ray scattering (SAXS): SAXS measurements have been used to calculate the scattering from the hydrogels. We used the Gauss-Lorentz Gel Model that is assumed to be valid for gel structures [38] that modeled the scattered intensity (Iq) as a sum of a low-q exponential decay plus a lorentzian at higher q-values:

$$I(q) = I_G(0) \exp\left(-\frac{q^2 \Xi^2}{2}\right) + \frac{I_L(0)}{(1+q^2\xi^2)} \qquad (2)$$

Where $\Xi$ is the static correlation length in the gel, which can be attributed to the 'frozen-in' crosslinks of some gels, $\xi$ is the dynamic correlation length, related to the fluctuating polymer chain between crosslinks. $I_G(0)$ and $I_L(0)$ are the scaling factors for each of these structures.

Conductometric titrations: The pH of a TOC suspension (2 mgml$^{-1}$) was brought to 2.0 by HCl and 1mM NaCl was added. Then, the pH was raised until 13.0 by addition of small aliquots of sodium hydroxide (0.1 M). The pH and conductivity values were recorded after each addition by a HD 2256.2 (Delta Ohm (Padova, Italy) instrument.

**Acknowledgements**

This work was partially supported by the Italian Ministry of University and Research through the "Futuro in Ricerca" project RBFR12OO1G-NEMATIC.

# Role of sonication pre-treatment and cation valence in nano-cellulose suspensions sol-gel transition.

C. A. Maestri, M. Abrami, S. Hazan, E. Chistè, Y. Golan, J. Rohrer, A. Bernkop-Schnürch, M. Grassi, M. Scarpa, P. Bettotti

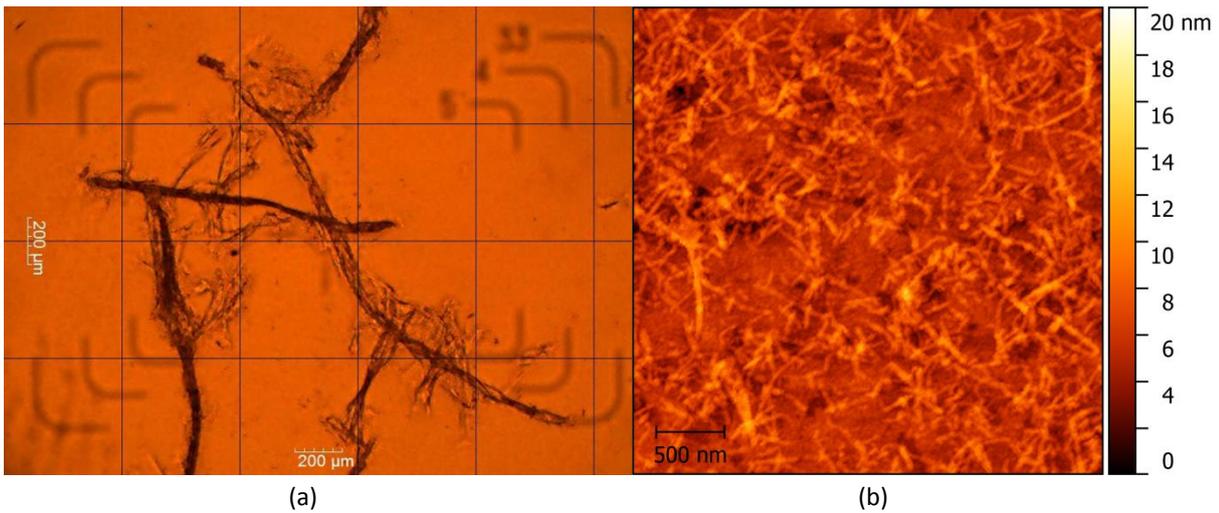

Figure S1. (a) Optical microscopy image of macroscopic cellulose fibers after a short sonication treatment (30 s); (b) Atomic force microscopy of cellulose nanocrystals after a prolonged sonication treatment (480 s).

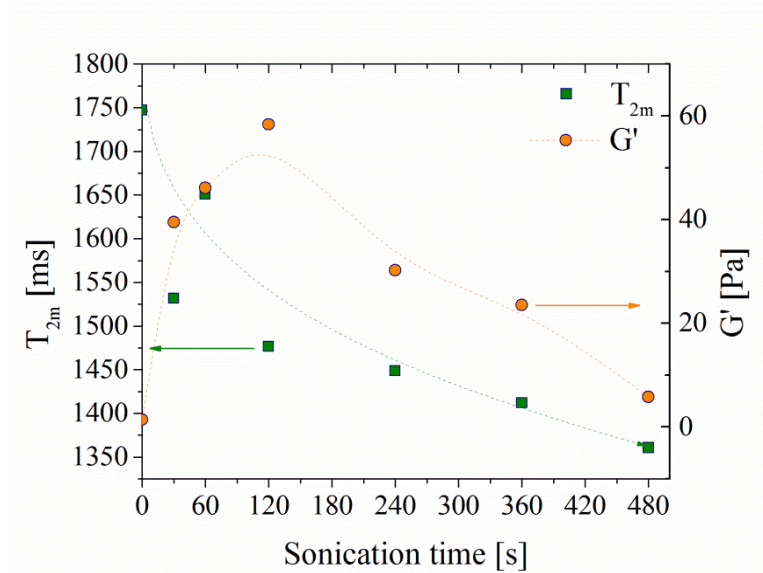

Figure S2. (Left axis) $T_{2m}$ data for samples without salt addition. A single decay time describes the magnetic relaxation. (Right axis) The G' modulus corresponding to different sonication times. The decrease of G' vs sonication time support the hypothesis that a more homogeneous dispersion of the fibres is less entangled and exposes a larger surface available to interact with water molecules (lines are guides for the eyes).

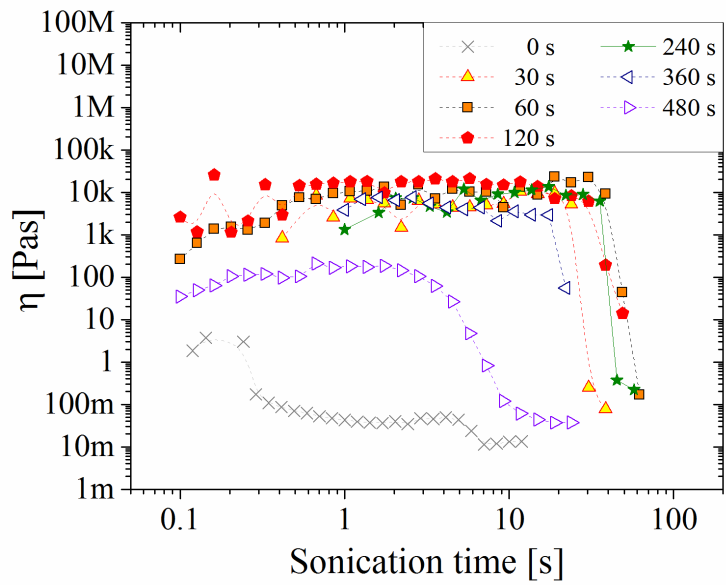

Figure S3. Flow curves for gels without salts. The system shows a dependence of its viscosity vs. sonication time. The maximum value of η is obtained for sonication time between 30 and 240 s. In all cases η shows a marked drop for shear stress above 30 Pa

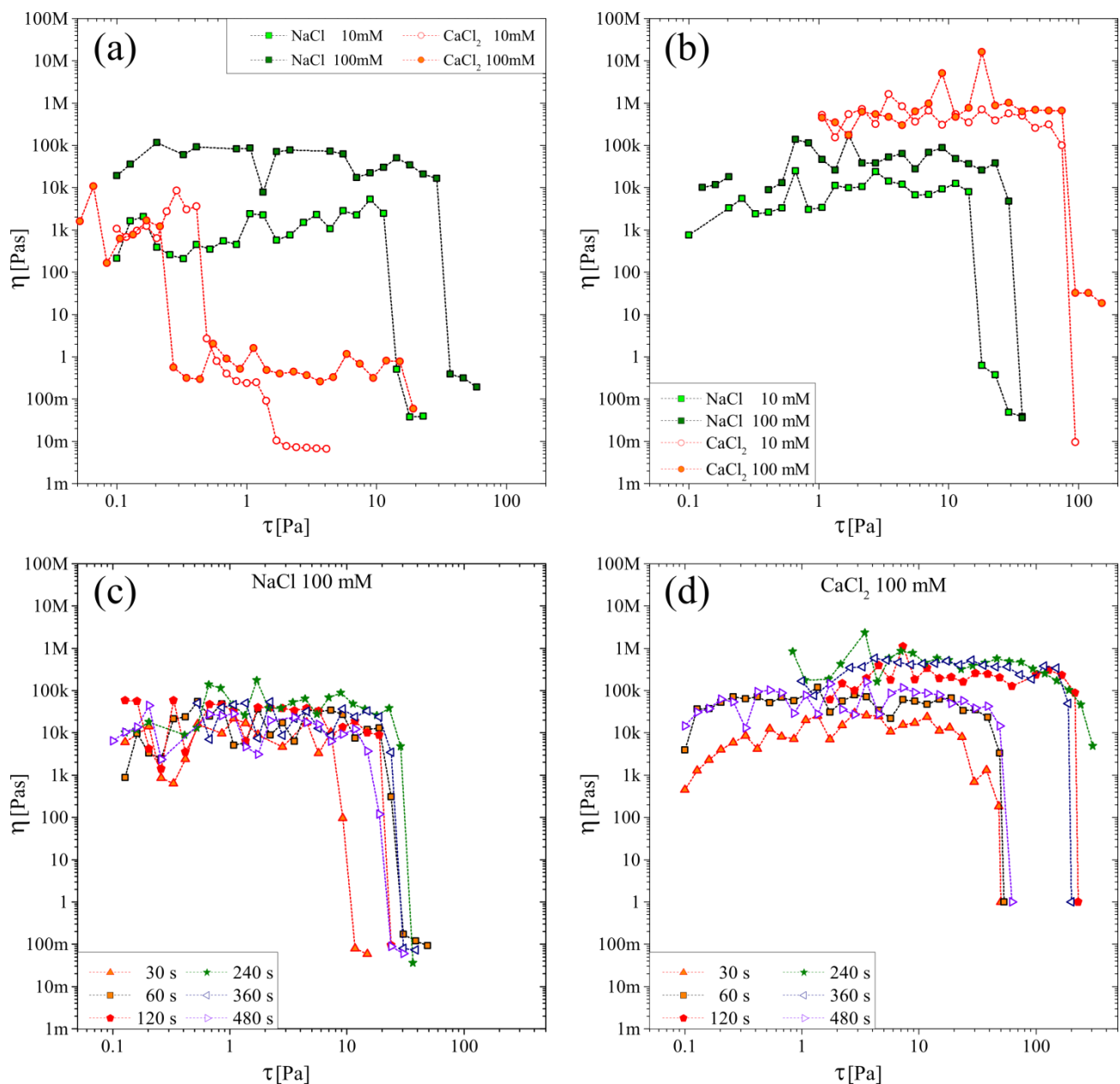

Figure S4. Flow curves for gels formed upon salts addition. (a) Addition of salts followed by sonication of the TOCs creates an inhomogeneous gel. The Na-based gels are homogeneous, compared to the Ca-based systems, and this fact reflects into a larger value of the viscosity drop, even if they show a smaller G' modulus. (b) If salts are added after sonication, homogeneous gel are formed and flow curves show a clear increase of the inflection point versus cation valence and concentration. (c) Gels formed by addition of 100 mM NaCl after sonication show an increase of threshold shear stress that determines the decrease of the viscosity for sonication times up to 240 s, then the threshold is slightly reduced for longer sonication treatment. (d) In the case of 100 mM $CaCl_2$, added after sonication, the threshold shear stress is, again, maximized for 240 s long sonication, but its decrease is more marked compared to NaCl case.

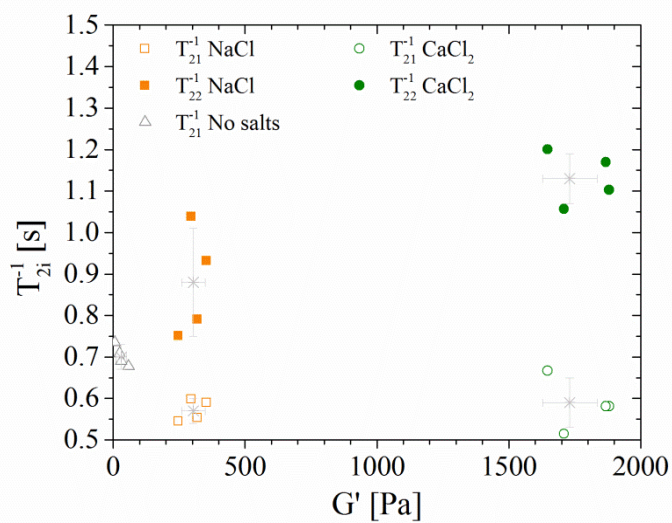

Figure S5. Correlation between $T_{2i}$ components and G' for both gels formed with and without salts. Only data for samples added with 100 mM salts and sonicated for at least 120 s are shown. Assuming the $T_{2i}$ components describes the local environment of the water molecules, the gel formed without salt addition show a $T_{21}$ relaxation rate intermediate compared to the two components needed to describe the magnetization relaxation in gels formed upon salt addition. Cation induced gels show a comparable $T_{21}$ rate for both NaCl and $CaCl_2$ samples, while the $T_{22}$ rate is significantly larger in $CaCl_2$ containing gels, indicating a different local structure of the water molecules that determines the greater G' modulus achieved in these gels.